# Using Machine Learning and Natural Language Processing to Review and Classify the Medical Literature on Cancer Susceptibility Genes

Yujia Bao[1*], M.A; Zhengyi Deng[2*], M.S.; Yan Wang[2], M.D.; Heeyoon Kim[1], M.S.; Victor Diego Armengol[2], A.B.; Francisco Acevedo[2], M.D.; Nofal Ouardaoui[3]; Cathy Wang[3,4], M.S.; Giovanni Parmigiani[3,4], Ph.D.; Regina Barzilay[1], Ph.D.; Danielle Braun[3,4], Ph.D.; Kevin S Hughes[2,5], M.D.

[1] Computer Science & Artificial Intelligence, Massachusetts Institute of Technology, Boston, MA;

[2] Department of Surgical Oncology, Massachusetts General Hospital, Boston, MA;

[3] Department of Biostatistics, Harvard T.H. Chan School of Public Health, Boston, MA;

[4] Department of Data Sciences, Dana-Farber Cancer Institute, Boston, MA;

[5] Harvard Medical School, Boston, MA

*These authors contributed equally to this study and should be considered as co-first author.

Acknowledgements of research support:

This work was supported by the National Cancer Institute at the National Institutes of Health (5T32CA009337-32, and 4P30CA006516-51) and the Koch Institute/Dana-Farber/Harvard Cancer Center Bridge Project (Footbridge).

Corresponding author:

Danielle Braun, PhD. Department of Biostatistics, Harvard T.H. Chan School of Public Health, Department of Data Sciences, Dana-Farber Cancer Institute, 677 Huntington Ave, SPH 4th Floor, Boston, MA 02115. Tel: +1-617-632-3654; Fax: +1-617-632-5444; E-mail: dbraun@mail.harvard.edu

Running head:

Using Natural Language Processing to Review Medical Literature

List of where this study has been presented:

- MGH's Clinical Research Day (October 5, 2017). Title: Using Machine Learning and Natural Language Processing to Predict Cancer Risk after Multi-Gene Panel Testing.
- DF/HCC Junior Investigator Symposium (November 6, 2017). Title: Using Machine Learning and Natural Language Processing to Predict Cancer Risk after Multi-Gene Panel Testing.
- Biostatistics and Computational Biology Annual Retreat DFCI (January 2018). Title: Using Machine Learning and Natural Language Processing to Predict Cancer Risk after Multi-Gene Panel Testing.
- 2018 DFCI/FSTRF Marvin Zelen Memorial Symposium (April 6, 2018). Title: Using Machine Learning and Natural Language Processing to Predict Cancer Risk after Multi-Gene Panel Testing.

Disclaimers: Dr. Hughes receives Honoraria from Myriad Genetics Veritas Genetics, Advisory Board for Beacon (An RFID Biopsy Marker) and is a founder of and has a financial interest in Hughes Risk Apps, LLC. Dr. Hughes's interests were reviewed and are managed by Massachusetts General Hospital and Partners Health Care in accordance with their conflict of interest policies.

Dr. Parmigiani is a member of the Scientific Advisory Board and has a financial interest in Cancer Risk Apps LLC (CRA). CRA commercializes software for management of patients at high risk of cancer. He is also a member of the Scientific Advisory Board of Konica-Minolta who owns Ambry genetics.

We feel there is no significant overlap with this work.

# ABSTRACT

**PURPOSE**: The medical literature relevant to germline genetics is growing exponentially. Clinicians need tools monitoring and prioritizing the literature to understand the clinical implications of the pathogenic genetic variants. We developed and evaluated two machine learning models to classify abstracts as relevant to the penetrance (risk of cancer for germline mutation carriers) or prevalence of germline genetic mutations.

**METHODS**: We conducted literature searches in PubMed and retrieved paper titles and abstracts to create an annotated dataset for training and evaluating the two machine learning classification models. Our first model is a support vector machine (SVM) which learns a linear decision rule based on the bag-of-ngrams representation of each title and abstract. Our second model is a convolutional neural network (CNN) which learns a complex nonlinear decision rule based on the raw title and abstract. We evaluated the performance of the two models on the classification of papers as relevant to penetrance or prevalence.

**RESULTS**: For penetrance classification, we annotated 3740 paper titles and abstracts and used 60% for training the model, 20% for tuning the model, and 20% for evaluating the model. The SVM model achieves 89.53% accuracy (percentage of papers that were correctly classified) while the CNN model achieves 88.95 % accuracy. For prevalence classification, we annotated 3753 paper titles and abstracts. The SVM model achieves 89.14% accuracy while the CNN model achieves 89.13 % accuracy.

**CONCLUSION**: Our models achieve high accuracy in classifying abstracts as relevant to penetrance or prevalence. By facilitating literature review, this tool could help clinicians and researchers keep abreast of the burgeoning knowledge of gene-cancer associations and keep the knowledge bases for clinical decision support tools up to date.

## INTRODUCTION

The medical literature is growing exponentially, and nowhere is this more apparent than in genetics. In 2010, a PubMed search for "BRCA1" yielded 7,867 papers, while in 2017 the same search retrieved nearly double that amount (14,266 papers). As the literature about individual genes increases, so does the number of pathogenic gene variants that are clinically actionable. Panel testing for hereditary cancer susceptibility genes identifies many patients with pathogenic variants in genes that are less familiar to clinicians, and it is not feasible for clinicians to understand the clinical implications of these pathogenic variants by conducting their own comprehensive literature review. Thus, clinicians need help monitoring, collating and prioritizing the medical literature. In addition, clinicians need clinical decision support tools to help facilitate decision-making for patients. These tools depend on a knowledge base of the metadata on these genetic mutations that is both up-to-date and comprehensive.[1]

Natural language processing (NLP) is an area of Artificial Intelligence (AI) that focuses on problems involving the interpretation and "understanding" of free text by a non-human system.[2,3] Traditional NLP approaches have relied almost exclusively on rules-based systems, where domain experts predefine a set of rules used to identify text with specific content. However, defining these rules is laborious and challenging as a result of variations in language, format and syntax.[4] Modern NLP approaches instead rely on machine learning, where predictive models are learned directly from a set of texts that have been annotated for the specific target.

NLP has been applied in fields relevant to medical and health research.[2,5,6] For example, in the field of oncology, researchers have used NLP to identify and classify cancer patients, assign staging, and determine cancer recurrence.[7–9] NLP also has an important role in accelerating literature review by classifying papers as relevant to the topic of interest.[10,11] Several studies developed and improved machine learning approaches based on the publicly available literature collections of 15 systematic literature reviews.[11–14] These reviews were conducted by the Evidence-based Practice Centers to evaluate the efficacy of medications in 15 drug categories.[13] Frunza et al. used a complement naïve Bayes approach to identify papers on the topic of the dissemination strategy of health care services for elderly people and achieved a precision of 63%.[15] Fiszman et al. proposed an approach to identify papers relevant to cardiovascular risk factors (56% recall, 91% precision).[16]

Miwa et al. extended an existing approach to classify social and public health literature on topics of cooking skills, sanitation, tobacco packaging, and youth development.[17]

However, no NLP approaches have been developed specifically for classifying literature regarding the penetrance (risk of cancer for germline mutation carriers) or prevalence of germline genetic mutations. To our knowledge, no annotated dataset is available for the purpose of developing a machine learning method to identify relevant papers in this domain. In this study, we aimed to create a human-annotated dataset of abstracts on cancer susceptibility genes and develop a machine learning-based NLP approach to classify abstracts as relevant to the penetrance or prevalence of pathogenic genetic mutations.

**MATERIALS AND METHODS**

Institutional Review Board approval was not needed as no human data were analyzed.

**Establishing an annotated dataset**

To develop effective machine learning models for automatic identification of relevant papers, we created a human-annotated dataset. We performed PubMed searches using the following query templates:

- *("gene name"[TIAB] OR "medical subject headings (MeSH) for that gene" OR "related syndrome name"[TIAB] OR "MeSH for that syndrome") AND ("Risk"[Mesh] OR "Risk"[TI] OR "Penetrance"[TIAB] OR "Hazard ratio"[TIAB]) AND ("cancer name"[Mesh] OR "cancer name"[TIAB])*

- *("gene name"[TIAB] OR "medical subject headings (MeSH) for that gene" OR "related syndrome name"[TIAB] OR "MeSH for that syndrome").*

We considered different gene-cancer combinations from the All Syndrome Known to Man Evaluator (ASK2ME),[18] a recently developed clinical decision support tool for clinicians to estimate the age-specific cancer risk of germline mutation carriers. This tool captures most of the important gene-cancer combinations. We opted to use the title and abstract of each paper as the input for our models, for three main reasons. First, this information can be automatically downloaded through EDirect[19], whereas automatically downloading the full-text papers was not feasible due to licensing issues. Second, the title and abstract of each paper can be downloaded in free text form, whereas full-text papers are not generally available in a common format, and one needs to handle PDF, HTML, as well as others. Last but not least, annotating the title and abstract is less time-consuming than annotating the full-text, and therefore obtaining a large training dataset is feasible. Each paper (based on title and abstract) was annotated for the following fields by six human annotators, with a minimum of two human annotators per paper. Two fields (penetrance and prevalence) were used to classify papers as relevant to penetrance, prevalence, both, or neither. Other fields (polymorphism, ambiguous penetrance, ambiguous incidence) were annotated and used as exclusion criteria.

- Penetrance: presence of information about risk of cancer for germline mutation carriers.
- Prevalence: presence of information about proportion of germline mutation carriers in the general population or among individuals with cancer.
- Polymorphism: presence of information only on a germline genetic variant present in more than 1% of the general population.
- Ambiguous penetrance: a) unresolved disagreement between human annotators on the penetrance label or b) impossibility of determining the penetrance label solely based on the title and the abstract.
- Ambiguous prevalence: a) unresolved disagreement between human annotators on the prevalence label or b) impossibility of determining the prevalence label solely based on the title and the abstract.

Our goal was to develop models that could accurately classify papers with subject matter pertaining to the penetrance and prevalence of rare germline mutations. Papers annotated as polymorphism or ambiguous were not used for model training, tuning or evaluation.

**Models**

Our first model is a support vector machine (SVM). We first tokenized the input title and abstract and converted them into a standard bag-of-ngram vector representation. Specifically, we represented each title and abstract by a vector, wherein each entry is the term frequency-inverse document frequency (tf-idf) of the corresponding ngram. Tf-idf increases in proportion to the frequency of the ngram in this particular abstract and is offset by the frequency of the ngram in the entire dataset. Thus, the resulting representation serves to down-weight the feature value of common words that add little information, such as articles. Finally, we used this bag-of-ngram representation as the input for a linear SVMs to predict its corresponding label.

Our second model is a convolutional neural network (CNN).[20] This model directly takes the tokenized title and abstract as its input and convolves the input with learnable local filters. It then combines the convolved representation globally and predicts the final label. Unlike the linear SVM, the CNN model is capable of learning nonlinear decision rules.

**Model Evaluation**

For both the penetrance and the prevalence classification task, we split the dataset randomly into a model training set (60% of the data), a model tuning set (20% of the data), and a model evaluation (20% of the data). We trained the two models on the training set and used the model tuning set for hyper-parameter selection. The model performance was evaluated on the model evaluation set. We used accuracy (percentage of the papers that were correctly classified) and F1 score as our evaluation metrics. Here, the F1 score is the harmonic mean of precision (percent of predicted positive that are true positive) and recall (percent of all true positives that are predicted as positive). Learning curves were constructed showing how the number of papers annotated in the training set affects the accuracy of the models. We also plotted the receiver operating characteristic curve (ROC curve) to compare the model performance at various thresholds.

**RESULTS**

**Dataset**

The final human-annotated dataset contained 3,919 annotated papers **(Table 1)**. Of these, 989 were on penetrance and 1291 were on prevalence. We excluded papers that were labeled as polymorphism-related. For the task of penetrance classification, we further excluded papers with an ambiguous penetrance label, reducing the annotated dataset to 3740. For the task of prevalence classification, we excluded papers with ambiguous prevalence label, reducing the annotated dataset to 3753 **(Table 1)**.

**Model Performance**

Table 2 shows the performance of the SVM and CNN model. The SVM model achieves 0.8953 accuracy and 0.7886 F1 score in penetrance classification and 0.8914 accuracy and 0.8396 F1 score in prevalence classification. Although the CNN has more flexibility in modeling, it underperforms by a small margin compared to the SVM model. **Figures 1a** and **1b** show the receiver operating characteristic curve (ROC curve) of the two models, for penetrance and prevalence classification respectively. The y-axis is the true positive rate, which is also known as sensitivity or recall. The x-axis is the false positive rate, which represents the probability of false alarm. The ROC curve provides a comparison of the model performance at different levels of decision threshold. Both models achieved similar area under the ROC curve (AUC) for both classification tasks. **Figures 2a** and **2b** depict the learning curves for the two models for penetrance and prevalence classification respectively. For penetrance classification, when only 500 annotated papers are used for training, the SVM model achieved around 0.89 accuracy, while the CNN model achieves less than 0.85 accuracy. However, the learning curve of the CNN model improves steadily as the training set increases. For prevalence classification, the two learning curves show a flattening trend after the number of papers reaches 1000.

**DISCUSSION**

The growing number of cancer susceptibility genes identified and the burgeoning literature regarding these genes is overwhelming for clinicians and even for researchers. Machine learning algorithms can help identify the relevant literature. In this study, we have created a dataset containing almost four thousand human annotated papers regarding cancer susceptibility genes. Using this dataset, we developed two models to classify papers as relevant to the penetrance or prevalence of cancer susceptibility genes. The SVM model we developed achieves 89.53% accuracy for penetrance and 89.14% accuracy for prevalence, outperforming the more complex CNN model. As we have shown in **Figures 2a** and **2b**, our models perform better as the number of papers in the training set increases. Although the curves will plateau at some point, the increasing trend indicates that model performance will continue to improve as more annotated papers are added to the training set.

To maximize efficiency, SVM-based NLP approaches have been developed to identify relevant papers in the medical literature for various topics. In 2005, Aphinyanaphongs et al. developed the first SVM method to assist systematic literature review by identifying relevant papers in the domain of internal medicine.[21] Several similar approaches were subsequently proposed, including an approach developed by Wallace et al. that incorporates active learning to reduce annotation cost.[11,22] Wallace et al. reduced the number of papers that must be reviewed manually by around 50%, while capturing all important papers for systematic review.[22] Fiszman et al. developed a system using symbolic relevance processing to identify potentially relevant papers for cardiovascular risk factor guidelines. The performance of his system was 56% recall and 91% precision.[11,16] While most existing methods have focused on the clinical literature, recently Miwa et al. extended the scope of their approach to include the social science literature.[17] CNN-based NLP methods have been developed for short text and sentence classification.[23–26] However, few methods have been developed and tested on classifying medical literature. Using the Risk of Bias (RoS) text classification datasets, Zhang et al. developed a CNN model to assess the study design bias in literature on randomized clinical trials (RCTs). The accuracy of the model ranged from 64% to 75%.[27]

The high accuracy and F1 score of the models we developed show that these models can be used to classify prevalence and penetrance papers regarding cancer susceptibility genes. This approach will be a useful tool for physicians to prioritize literature and understand the clinical implications of pathogenic variants. Also, this NLP approach has the potential to assist systematic literature review and meta-analysis in the same domain. We have conducted another study to test its efficiency and comprehensiveness in identifying important papers for meta-analyses which will be reported separately (Deng Z, Yin K, Bao Y, et al. Validation of a Semi-automated Natural Language Processing-based Procedure for Meta-Analysis of Cancer Susceptibility Gene Penetrance. Submitted to *JCO Clin Cancer Informatics*).

Although our approach achieves high performance, there are some limitations. One weakness of our approach is the dependence on data available in the title and abstract. This is partly due to limitations in access to full-text publications, but also due to the variety of formats in which full-text publications are stored. The proposed models do not work for papers that do not have an abstract or have an incomplete abstract. When the abstract is ambiguous for humans, misclassification can also occur. In the annotated training dataset, there are 119 papers (3.0%) having ambiguous penetrance information, and 101 papers (2.6%) having ambiguous prevalence information. Although we excluded these from model training, classifying new abstracts that are ambiguous remains challenging.

The abstract is an important component of a published work, and is usually available publicly. A well-written and complete abstract provides concise yet critical information pertinent to the study, can facilitate the capture of the key content by the reader, and can greatly facilitate NLP. When abstracts are not clearly written or leave out critical findings of the study, the efficacy of NLP models based on abstract text decreases. There is a need for authors to report their findings in sufficient detail if NLP methods are to be effective in the future.

One approach to handle important studies that do not have an abstract or do not report sufficient detail in the abstract is to develop classification algorithms based on the full text. Usually, full texts provide much more information on penetrance and prevalence. Developing algorithms to extract and read information from full texts may ultimately lead to higher accuracy. However, numerous issues will have to be solved to develop algorithms based on full text, including: (1) retrieving the PDF files of numerous papers automatically (including resolving access issues), (2) automatically extracting text, figures, and tables from a PDF or other published format, and (3) developing more complex classification models for additional labels.

As we have shown, the CNN model did not outperform the SVM model. This is true for both classification tasks and is not surprising, as neural networks typically require much larger amounts of annotated data for training. As an alternative to annotating more data, one may further improve model performance by asking human annotators to provide justifications for their decisions[28]. These justifications can be in the form of highlighting parts of the original input abstract that informed the classification decision. Recently, Bao et al. and Zhang et al. showed that providing these justifications to the model can significantly improve classification performance when limited amount of training data are available.[27,29]

In this study, we developed two models to classify abstracts relevant to the penetrance or prevalence of cancer susceptibility genes. Our models achieve high performance and have the potential to reduce the literature review burden. With the exponential growth of the medical literature, our hope is to use computing power to help clinicians and researchers search for and prioritize knowledge in this field and to keep knowledge bases that are used by Clinical Decision Support tools, such as ASK2ME[1,18], up to date.

**FIGURE LEGENDS**

Figure 1. Receiver operating characteristic curve for **(a)** penetrance classification, and **(b)** prevalence classification.

Figure 2. Learning rate of the two models on the task of **(a)** penetrance classification, and **(b)** prevalence classification.

# FIGURES

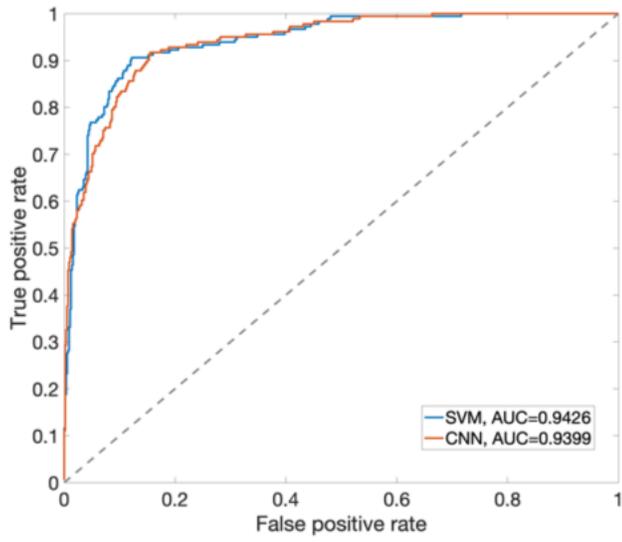 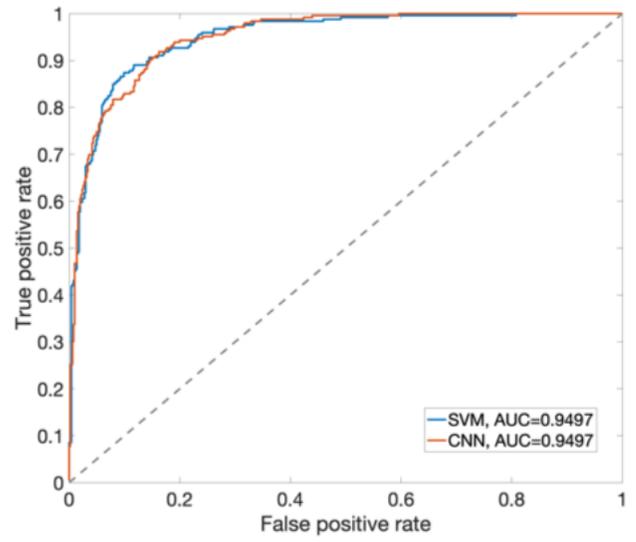

(a)        (b)

**Figure 1**

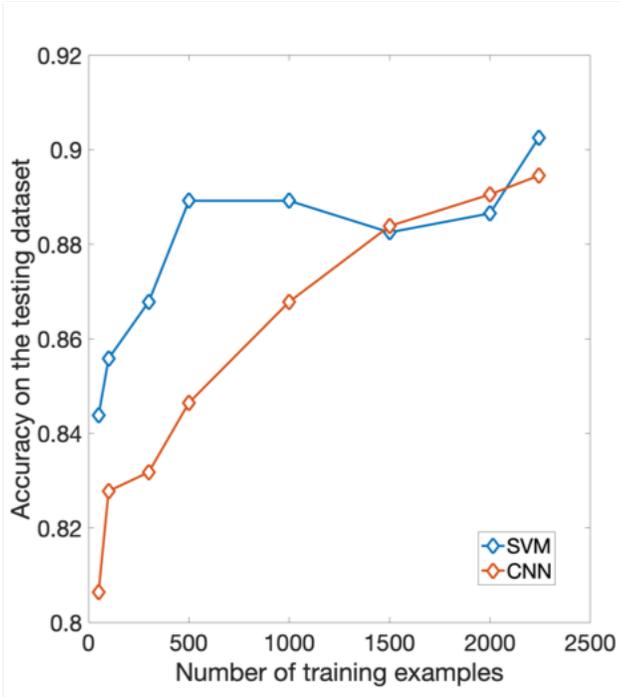 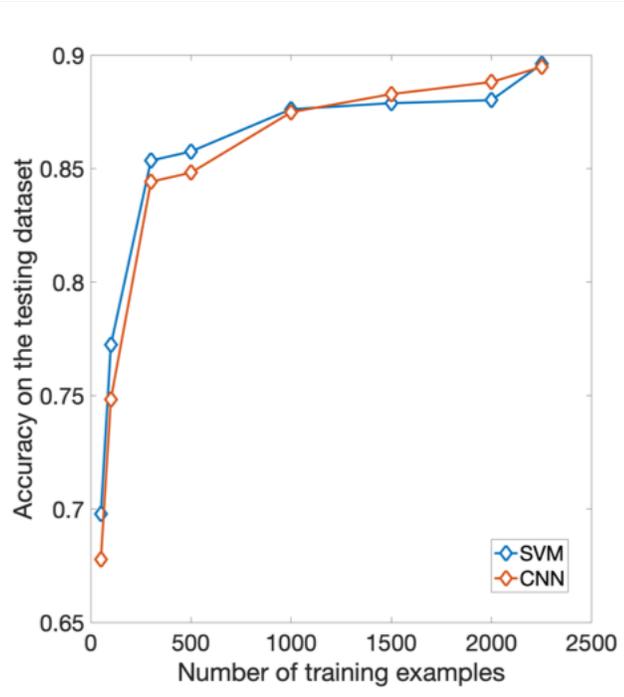

(a)             (b)

**Figure 2**

# TABLES

**Table 1. Summary of the annotated dataset**

|  | Number of Positive Papers | Number of Negative Papers |
|---|---|---|
| *Original Dataset* | | |
| Penetrance | 989 | 2930 |
| Prevalence | 1291 | 2628 |
| Polymorphism | 295 | 3624 |
| Ambiguous penetrance | 119 | 3800 |
| Ambiguous prevalence | 101 | 3818 |
| *After Excluding Polymorphism and Ambiguous Papers* | | |
| Penetrance | 904 | 2836 |
| Prevalence | 1230 | 2523 |

**Table 2. Performance of two NLP models developed for penetrance and prevalence classification**

| Task | Penetrance Classification | | Prevalence Classification | |
|---|---|---|---|---|
|  | Accuracy (95%CI) | F1 Score (95%CI) | Accuracy (95%CI) | F1 Score (95%CI) |
| SVM | 0.8953 (0.8919, 0.8987) | 0.7886 (0.7779, 0.7993) | 0.8914 (0.8863, 0.8965) | 0.8396 (0.8334, 0.8458) |
| CNN | 0.8895 (0.8855, 0.8935) | 0.7544 (0.7368, 0.7720) | 0.8913 (0.8859, 0.8967) | 0.8348 (0.8281, 0.8415) |

CI: confidence interval. SVM: support vector machine. CNN: convolutional neural network